# The Critical Role of Substrate in Stabilizing Phosphorene Nanoflake: A Theoretical Exploration

Junfeng Gao, Gang Zhang, Yong-Wei Zhang

Institute of High Performance Computing, A*STAR, Singapore 138632

**ABSTRACT**: Phosphorene, a new two-dimensional (2D) semiconductor, has received much interest due to its robust direct band gap and high charge mobility. Currently, however, phosphorene can only be produced by mechanical or liquid exfoliation, and it is still a significant challenge to directly epitaxially grow phosphorene, which greatly hinders its mass production and thus applications. In epitaxial growth, the stability of nanoscale cluster or flake on a substrate is crucial. Here, we perform *ab initio* energy optimizations and molecular dynamics simulations to explore the critical role of substrate on the stability of a representative phosphorene flake. Our calculations show that the stability of the phosphorene nanoflake is strongly dependent on the interaction strength between the nanoflake and substrate. Specifically, the strong interaction (0.75 eV/P atom) with Cu(111) substrate breaks up the phosphorene nanoflake, while the weak interaction (0.063 eV/P atom) with h-BN substrate fails to stabilize its 2D structure. Remarkably, we find that a substrate with a moderate interaction (about 0.35 eV/P atom) is able to stabilize the 2D characteristics of the nanoflake on a realistic time scale. Our findings here provide useful guidelines for searching suitable substrates for the directly epitaxial growth of phosphorene.

## 1. Introduction

Stimulated by many miraculous properties of graphene, other two-dimensional (2D) materials, such as h-BN[1, 2], transition metal dichalcogenides (TMDs)[3, 4], silicene[5-8] and phosphorene[9, 10], have also been produced and studied. Among those 2D materials, phosphorene, which was mechanically exfoliated in 2014,[9-11] is the latest member joining the 2D materials family. Similar to graphene, phosphorene has a honeycomb network; but unlike graphene, it is a direct band gap semiconductor possessing a rectangular unit cell and a puckered structure.[12] Due to this unique puckered feature, its mechanical[13], thermal[14] and electronic[12, 15-17] properties are highly anisotropic. For example, based on first-principles calculations, the hole mobility along the armchair direction (~640–700 $cm^2V^{-1}s^{-1}$) was predicted to be 16–38 times lower than that along the zigzag direction (~2×10$^4$ $cm^2V^{-1}s^{-1}$).[12] Remarkably, the electronic properties of phosphorene can be effectively tuned by various methods, such as layer number engineering[18-20], surface doping[16, 21], applying external strain[22, 23] and applying electrical field[24, 25]. For example, the electronic properties of phosphorene were found to be strongly layer-dependent: The band-gap can be tuned from ~2.0 eV[18, 26] (monolayer) to ~0.3 eV[9, 12, 18, 20] (bulk), and the work function can be changed from 5.16 V (monolayer) to 4.50 V (five layers)[19]. Kim *et al.* observed that the band-gap of few-layer phosphorene could be continuously reduced with the increase in dopant density, and at a critical dopant density, the semiconducting phosphorene became a Dirac semimetal, in which the energy vs. wave vector relation was linear along the armchair direction and quadratic along the zigzag direction.[16] By applying compression vertically to its surface, Stewart *et al.* predicted that the room-temperature electron mobility of bilayer phosphorene was increased to 7×10$^4$ $cm^2V^{-1}s^{-1}$ along the armchair direction.[22] Hence, compared with graphene and other 2D materials, phosphorene possesses many unique properties, such as a moderate band gap between graphene (0 eV)[27] and MoS$_2$ (2.0 eV)[28], a relatively high carrier mobility[12, 22], strongly anisotropic properties[9-14] and highly tunable electronic properties[7,15-21]. Therefore, phosphorene has many unique advantages in electronic and optoelectronic device applications.[10, 29]

Epitaxial growth techniques, such as chemical vapor deposition (CVD) and physical vapor deposition (PVD), have been widely used to grow 2D materials, including graphene[30-34], h-BN[1, 2], $MoS_2$[35, 36] and silicene[5, 7]. In fact, epitaxial growth is considered to be one of the most efficient methods to mass-produce 2D materials with high quality, large size and low cost.[2, 30, 33, 34, 37] Currently, however, phosphorene can only be produced by mechanical exfoliation[9-11] or liquid exfoliation[38, 39] from prepared bulk black phosphorus. It is still a significant challenge to directly grow monolayer or few-layer phosphorene by using epitaxial growth, which greatly hinders its mass production and thus applications. With regards to the successful epitaxial growth of graphene and other 2D materials, an important question arises: Can monolayer phosphorene be directly grown by epitaxial technique? Clearly, answer to this question is not only of significant scientific interest, but also of great impact to the applications of phosphorene.

During the growth of 2D materials, the existence of stable crystalline nanoflake is vitally important for its continuous growth. One of the important factors that influence the stability of a nanoflake is its interaction with substrate.[40-42] For example, on Rh(111)[6] and Ru(0001)[43] surfaces, the crystalline structure of silicene was found to be broken, while on Ag(111) surface, the substrate was able to stabilize the configurations of initial silicene nanoflake and thus promote the growth of silicene monolayer.[5-8] Considering its buckled, highly anisotropic and relatively less stable structure[12, 18-20], we can infer that the growth of phosphorene should be very sensitive to substrate. To the best of our knowledge, there is no report on the stability of phosphorene nanoflake on substrate. Then an important question is: What kinds of substrates are potentially suitable for the direct epitaxial growth of phosphorene? Moreover, in addition to the critical role in growth of phosphorene, the interactions of phosphorene with substrates are also relevant to a variety of interesting and important research topics and applications, such as surface chemical reactivity and kinetics, photocatalysis, interfacial charge separation, electronic band structure and charge transfer, *etc*.

In this work, we explore the stability of a representative phosphorene nanoflake containing 27 atoms (black $P_{27}$) on different substrates via both *ab initio* energy optimization and *ab initio* molecular dynamics (AIMD) simulations, with aim to search for suitable substrates for the direct epitaxial growth of phosphorene. The reasons that we choose the $P_{27}$ nanoflake are: 1) it possesses the characteristic 2D structure of phosphorene with the $C_2$ symmetry, 2) such a flake size has been found by many previous theoretical and experimental explorations to be important for the growth of 2D materials, such as graphene and silicene,[6, 31, 41, 44] and 3) it also possesses important edge features. The formation energy of a 2D cluster can be approximately calculated as the sum of the edges energies, $\sim \frac{1}{2}\sum_n l_n \gamma_n$ ($l_n$ is the edge length and $\gamma_n$ is the edge energy per length)[45]. As the edge energy of phosphorene is about 0.2~0.3 eV/Å,[46, 47] which is comparable to or even lower than that of graphene edge (about 0.3~1 eV/Å)[48-50], the thermal and dynamic stability of phosphorene edges should also be important to the growth of phosphorene. Therefore, as a pioneering attempt, we use this flake in our study. With regards to substrates, we have chosen three different substrates: the h-BN substrate with vdW interaction, the Cu(111) surface with chemical adsorption, and the h-BN substrate with an artificially modified vdW interaction. For comparison, the structure and stability of the free-standing $P_{27}$ flake are also studied. Our calculations show that the stability of the phosphorene nanoflake is strongly dependent on the interaction strength between the flake and substrate. Importantly, we find that the characteristic 2D structure of the flake may be stabilized by choosing a proper substrate. This finding may provide a basis for searching suitable substrates for the direct epitaxial growth of phosphorene.

## 2. Computational Methodology

Various simulation methods, such as phase field method[51, 52], kinetic Monte Carlo (kMC) method[53, 54], and classical molecular dynamics method,[55] have been used to study the growth mechanism of 2D materials. However, the predictability of these methods is strongly dependent on material and physical parameters adopted. For example, free-energies for different phases and edge attachment/detachment rates are required for phase field method; dominant kinetic processes and their energy barriers are needed for kMC; and atomic interaction potential is demanded for classical molecular dynamics. For phosphorene, these parameters are largely unknown. *ab initio* simulations have been justified to be a powerful and accurate tool to explore the substrate interaction, stability and structure transition of nanoscale materials[56], and have been adopted successfully to study the growth mechanism of 2D materials, including graphene, h-BN and silicene [40, 41, 48, 57]. Therefore *ab initio* energy optimizations and *ab initio* molecular dynamics simulations are adopted in this work.

All the *ab initio* calculations in this paper were performed by using density functional theory and plane wave pseudopotential technique with spin-polarization, as implemented in the Vienna Ab-initio Simulation Package (VASP)[58, 59]. Generalized gradient approximation (GGA) with the Perdew–Burke–Ernzerhof (PBE) functional was used to describe the exchange-correlation interaction.[60] Grimme's DFT-D2 approach was used to describe the vdW interaction.[61] Projector-augmented wave (PAW) method[62] was used to describe the core electrons. A plane-wave basis kinetic energy cutoff of 400 eV and a convergence criterion of $10^{-5}$ eV were used in all the calculations. A conjugate-gradient algorithm was used to relax the ions until the force was less than 0.02 eV/Å. PREC was set to Normal for both structural relaxation and *ab initio* molecular dynamics calculations, but it was increased to High for charge differential density calculations. The charge differential densities and slices were plotted with VESTA soft package. For the transition state calculations, a climbing-image nudged elastic band (cNEB) method[63] was used to find saddle points and minimum energy paths with quick-min optimizer until the force was less than 0.03 eV/Å.

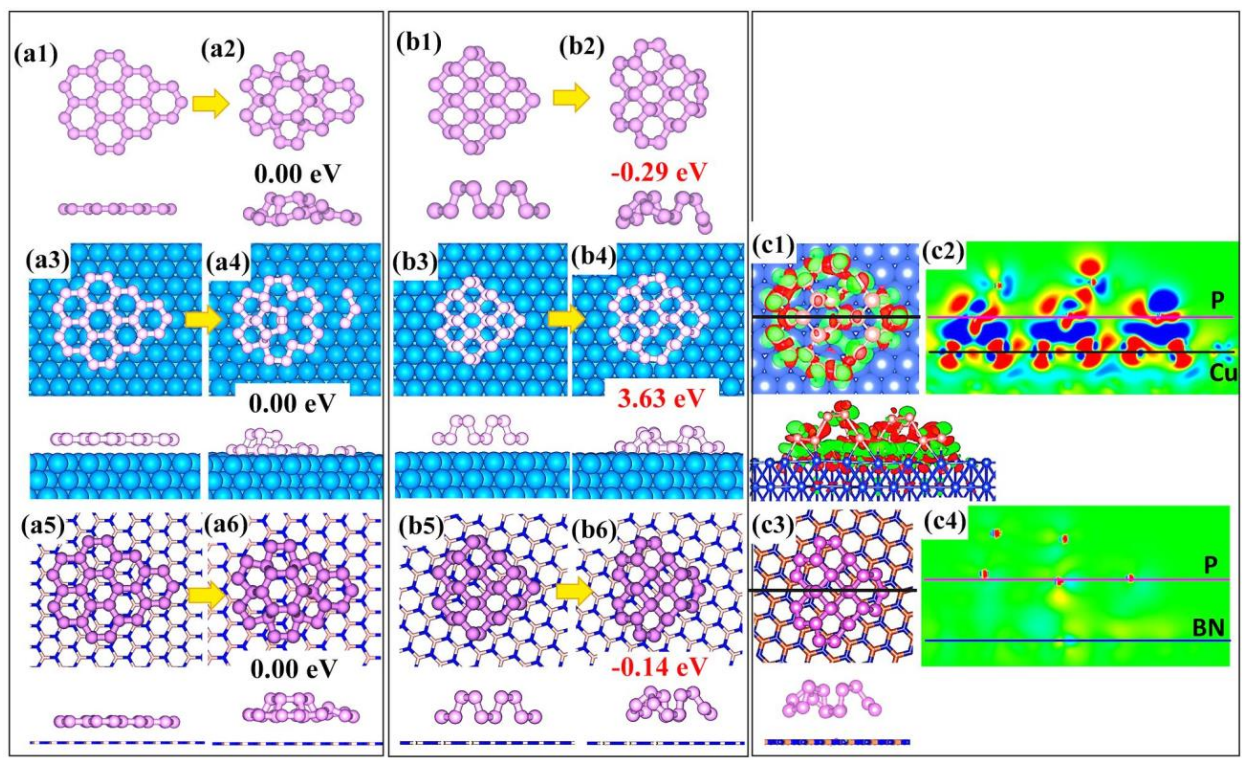

**Figure 1.** Left panel: Top view and side view of initially flat $P_{27}$ flake and optimized structures in vacuum (a1-2), on Cu(111) surface (a3-4), on h-BN surface (a5-6). Middle panel: Top view and side view of initially black $P_{27}$ and optimized structures in vacuum (b1-2), on Cu(111) surface (b3-4), on h-BN surface (b5-6), the red numbers are relative energies to the relaxed flat ones in left panel, respectively. Right panel: Charge differential densities of optimized black $P_{27}$ flake on Cu(111) (c1) and on h-BN (c3). Green (red) zone gains (loses) charges with isosurface value of $\pm 0.003|e|/bohr^3$, and the slices (c2, c4) are along the black lines in (c1, c3), respectively. The saturation level is $\pm 0.002|e|/bohr^2$.

## 3. Results and Discussion

### 3.1 Structure optimizations of phosphorene $P_{27}$ nanoflake

First of all, we checked the stability of $P_{27}$ flake in vacuum, on Cu(111) and h-BN surfaces, respectively, via *ab initio* energy minimizations. Here, we considered two types of $P_{27}$ nanoflake: the black $P_{27}$ with the phosphorene structure, and the flat $P_{27}$ with a perfectly planar structure. At the beginning, these two $P_{27}$ nanoflakes were placed in vacuum, on Cu(111) surface and h-BN surface (Left and middle panels of Figure 1), respectively. The h-BN substrate was chosen as a ($5\sqrt{3} \times 9$) monolayer supercell with dimensions of $(21.775 \times 22.626 \times 25.000)$ Å$^3$, while the Cu(111) substrate as a ($8 \times 5\sqrt{3}$) three-layer supercell with dimensions of $(20.448 \times 22.135 \times 25.000)$ Å$^3$. Such supercells are large enough so that the influences from the adjacent images are eliminated.

For the flat $P_{27}$ nanoflake, after the structural relaxation, the in-out buckling (Figure a2, a4, a6) arose spontaneously for all three scenarios[64], forming a two-layer structure resembling the phosphorene. Further examination showed that the relaxed $P_{27}$ nanoflake structure in vacuum (Figure a1-2) was nearly the same as that on the h-BN substrate (Figure a5-6). However, the relaxed $P_{27}$ nanoflake structure on the Cu(111) surface was found to break down spontaneously (Figure a3-4).

The stability of the black $P_{27}$ nanoflake in vacuum, on the Cu(111) surface and on the h-BN surface was also examined (Middle panel of Figure 1). Similar to the flat $P_{27}$ nanoflake, the black $P_{27}$ nanoflake in vacuum (Figure 1 (b1-2)) and on h-BN surface (Figure 1 (b5-6)) were similar and kept their 2D structural characteristic. On the Cu(111) surface, however, although most atoms reserved the 2D characteristic of phosphorene, one inner P-P bond was broken by forming P-Cu-P bonds (Figure 1 (b3-4)). Besides, two edge atoms in upper layer were pulled down to bottom by the strong interaction with the substrate. By comparing the relaxed flat $P_{27}$ nanoflake and relaxed black $P_{27}$ nanoflake, we found that the energy of the latter is 0.14 eV (0.29 eV) lower than the former on h-BN surface (in vacuum), implying the possibility of transforming the flat $P_{27}$ into the black $P_{27}$. In contrast, the energy of the black $P_{27}$ nanoflake is up to 3.63 eV higher than the broken one, indicating that the growth of phosphorene is difficult on the Cu(111) surface.

**Table 1.** The binding energy ($E_B$) (eV/P atom) of the relaxed black $P_{27}$ nanoflake on h-BN, Cu(111) and artificially modified h-BN surfaces. $E_B = (E_P + E_{Sub} - E_T)/N_P$, $E_T$ is the total energy of the black $P_{27}$ nanoflake on substrate, $E_P$ is the energy of the black $P_{27}$ nanoflake in vacuum, $E_{Sub}$ is the energy of related substrate, and $N_P$ is the number of P atoms, i.e. 27.

|       | h-BN  | Cu(111) | Modified h-BN |
|-------|-------|---------|---------------|
| $E_B$ | 0.063 | 0.754   | 0.349         |

The breakdown of the $P_{27}$ nanoflake on the Cu(111) is clearly due to the strong interaction between the nanoflake and Cu(111) surface. Although the Cu(111) surface has been used for the epitaxial growth of graphene, here we show that it is not suitable for the growth of phosphorene due to their strong interaction. Analysis on the charge differential densities for the relaxed black $P_{27}$ on the Cu(111) was shown in Figure 1 (c1-2). It is seen that the abundant charges were transferred into the zone between the $P_{27}$ and Cu(111) substrate both at the edge and inner part. Such scenario is similar to silicene on Ag(111) surface[6], but very different from graphene on metal surfaces[32, 41, 65]. When we consider the whole $P_{27}$ nanoflake, the calculated binding energy is about 0.754 eV/P atom (Table 1). But from the charge differential densities (Figure 1-c1) and its slice (Figure 1-c2), we can see that the binding is mainly attributed by 17 P atoms at the bottom layer. Hence, if we only consider the bottom layer P atoms, the binding energy is about 1.198 eV/P atom, signifying the strong interaction between the $P_{27}$ nanoflake and Cu(111) surface. In contrast, there is nearly no charge transfer between the $P_{27}$ and h-BN surface at the same isosurface level (Figure 1-c3&4) and the binding energy is only 0.063 eV/P atoms averaged over 27 P atoms.

Due to its chemical stability, h-BN was used as an effective capping layer to protect phosphorene from its structural and chemical degradation. Experimentally, phosphorene was found to be able to largely maintain its electronic and Raman characteristics in the h-BN/phosphorene heterostructure[66-69]. Theoretically, the interaction between BN and monolayer phosphorene was found to be vdW in nature [70, 71]. Therefore, the low binding energy between the $P_{27}$ and h-BN surface from the present calculation is consistent with those previously reported results.

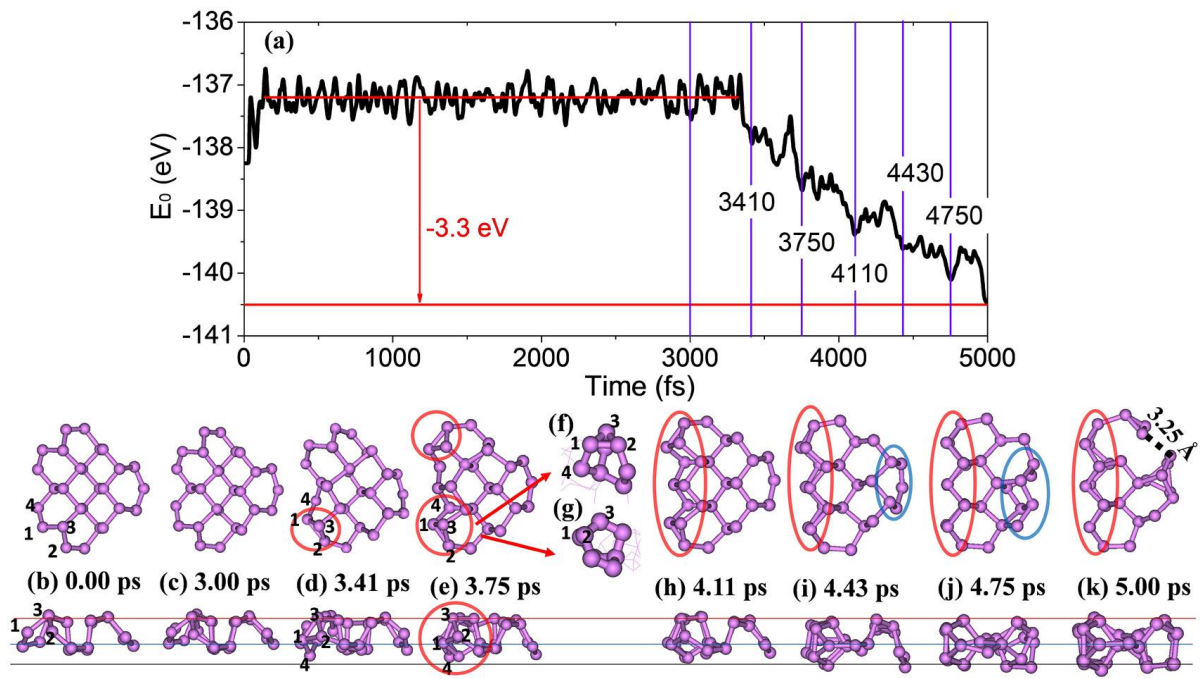

**Figure 2.** The free energy (a) of black $P_{27}$ nanoflake in vacuum during 5.0 ps AIMD simulations and their top view and side view of structural evolutions in time (b-e, h-k). Inserts (f) and (g) are the enlarged views of deformed zone in (e). Red and blue circles label the structure deformations. The lines in side views roughly divide the $P_{27}$ nanoflake into three layers.

## 3.2 Dynamic stability of phosphorene $P_{27}$ nanoflake in vacuum

Next, we explored the kinetic and dynamic stability of the black $P_{27}$ nanoflake via AIMD simulations. In the structure relaxation, we have shown that the Cu(111) surface is not a good candidate. Hence here, we focus on the AIMD simulations of the $P_{27}$ nanoflake in vacuum and h-BN substrate. The calculations were performed at 300 K (thermal degradation[72] temperature of few-layer black phosphorus is about 400 $^oC$). Each AIMD calculation lasted 5.0 ps and each step for ionic movement was 1.0 fs. For the AIMD simulations and the transition state calculations, a ($4\sqrt{3}\times7$) h-BN supercell with dimensions of (17.420 × 17.598 × 25.000) $\text{Å}^3$ was used, which is slightly smaller than the one used in the structure optimization. A canonical ensemble was adopted for the AIMD calculations using the algorithm of Nosé.[73] During the AIMD simulations, all boron and nitrogen atoms were fixed. As the melting point of h-BN is higher than 3000 K, such a freezing approach is reasonable.

Figure 2 showed the energetic and structural evolution of the black $P_{27}$ nanoflake in vacuum. It is seen that the black $P_{27}$ kept its structure only till ~3.4 ps (Figure 2a-c and Movie-S1&2 in Supporting Information (SI)). The first deformation took place by forming a P1-P2 bond between P1 and P2 atoms as labeled in Figure 2, and the neighboring P4 atom was pulled down from the initial bottom layer (Figure 2d). The driving force for the deformation was originated from the passivation of unsaturated P1 and P2 atoms at the edge, which reduced ~0.6 eV in energy.

Once the P1-P2 bond was formed, the subsequent structure changes followed quickly. Another edge P-P bonding occurred at the symmetric position of P1-P2 (Figure 2e) only in 0.34 ps later. From the enlarged view in Figure 2f&g, the structure change was similar to previously reported $P_8^-$ anions.[74] The subsequent structure changes were also easy, with an interval of every 0.3 ps. In Figure 2h&i, a new reconstructed edge was formed by downwarping the original edge and connecting secondary edge atoms (see the Movie-S2 in SI). The main structure of the newborn edge was composed of a series of deformed pentagons (Figure 2h), which are very similar to red phosphorus or previously reported $C_{2v}$–symmetric phosphorus chains, which was one of the most stable configurations in vacuum.[64] Therefore, such edge reconstruction may actually exist in phosphorene nanoribbons and quantum dots.

The downwarping and deformation of the opposite edge occurred at 4.43 ps (Figure 2i&j), which further reduced the energy. At about 5.0 ps, the strain introduced by the deformation finally broke the $P_{27}$ nanoflake (see Figure 2k). After the 5.0 ps of AIMD simulations, the two-layer black $P_{27}$ was transformed into a three-layer structure (see the dividing lines in the side view of Figure 2b-k), losing the 2D characteristics of phosphorene. Via these structure changes, the total energy was about 3.3 eV lower than that of the initial black $P_{27}$. Therefore it is clear that the black phosphorene nanoflake is not kinetically/dynamically stable in vacuum.

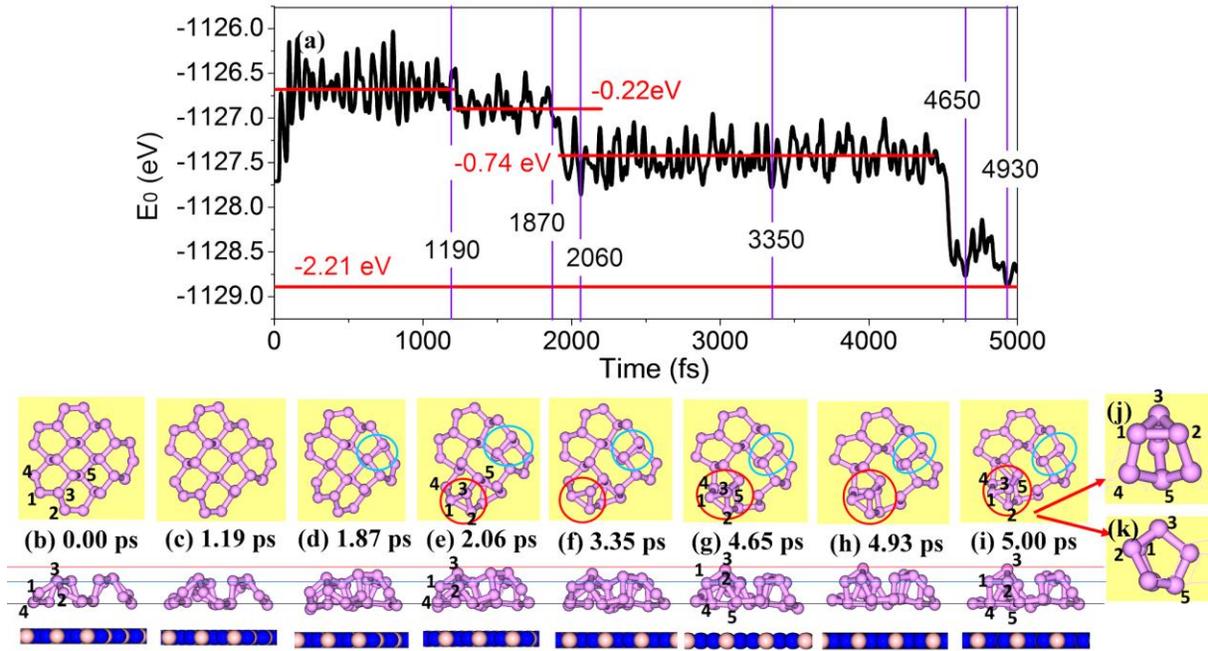

**Figure 3.** The free energy of the black $P_{27}$ nanoflake on the h-BN surface during 5.0 ps AIMD simulations (a) and the structural evolutions at different times (b-i). Inserts (j) and (k) are the enlarged views of the deformed zone in (i). Red and blue circles label the structure deformations. The yellow background in the top view represents the h-BN substrate for a clear view. The lines in the side views divide the $P_{27}$ nanoflake in three layers roughly.

**3.3 Dynamic stability of phosphorene nanoflake on h-BN**

The energetic and structural evolutions of the black $P_{27}$ nanoflake on the h-BN surface were shown in Figure 3. Although the interaction between black $P_{27}$ and h-BN surface is weak (Table 1 and Figure 1c3&4), the deformation of $P_{27}$ on the h-BN, which is step by step (Figure 3a and Moive-S3 in SI), is very different from the nearly continuous structure changes in vacuum. At the first stage, the initial black $P_{27}$ kept its structure till about 1.2 ps (Figure 3a-c). Then the protruded edge at the right part of $P_{27}$ shrank back and formed a new P-P bond just under the upper layer P-P bond in the blue circle of Figure 3d. This deformation reduced the energy by only ~0.22 eV, but it didn't change the two-layer characteristic of black phosphorene nanoflake (Figure 3a). A marked energy reduction of ~0.74 eV happened in the third stage of deformation, by forming a P1-P2 bond (red circle in Figure 3e), similar to that in vacuum (Figure 2d). But the P4 atom could not go down due to the confinement of the h-BN substrate, thereby P5 atom was pushed up from the upper layer. This deformation has a much longer lift time (> 2.5 ps, Figure 3e-g) than that in vacuum since the confinement of the h-BN substrate limited the edge reconstruction. Via the motion of P5 atom, the energy of the distorted configurations in Figure 3h&i was about 2.2 eV lower than the initial black $P_{27}$ nanoflake. As shown in the amplifying view (Figure 3j-k), the deformation looked like the previously reported anion $P_7^-$ cluster in vacuum.[74] The P3 atom was further lifted up to ~1 Å higher than the initial upper layer, signifying a new P layer was going to form (see the dividing line in side view of Figure 3b-i). Overall, the inner structure changes and energy reduction of the black $P_{27}$ on the h-BN surface were lower than that in vacuum (see the Moive-S3, S4 in SI), but they were still markedly visible. Although the h-BN substrate shows some positive effects on the stability of the $P_{27}$ nanoflake, the weak interaction with h-BN surface with ~0.063 eV/atom is unable to stabilize the 2D characteristics of phosphorene. Clearly, a stronger $P_{27}$-substrate interaction is required to stabilize the 2D characteristics of the nanoflake.

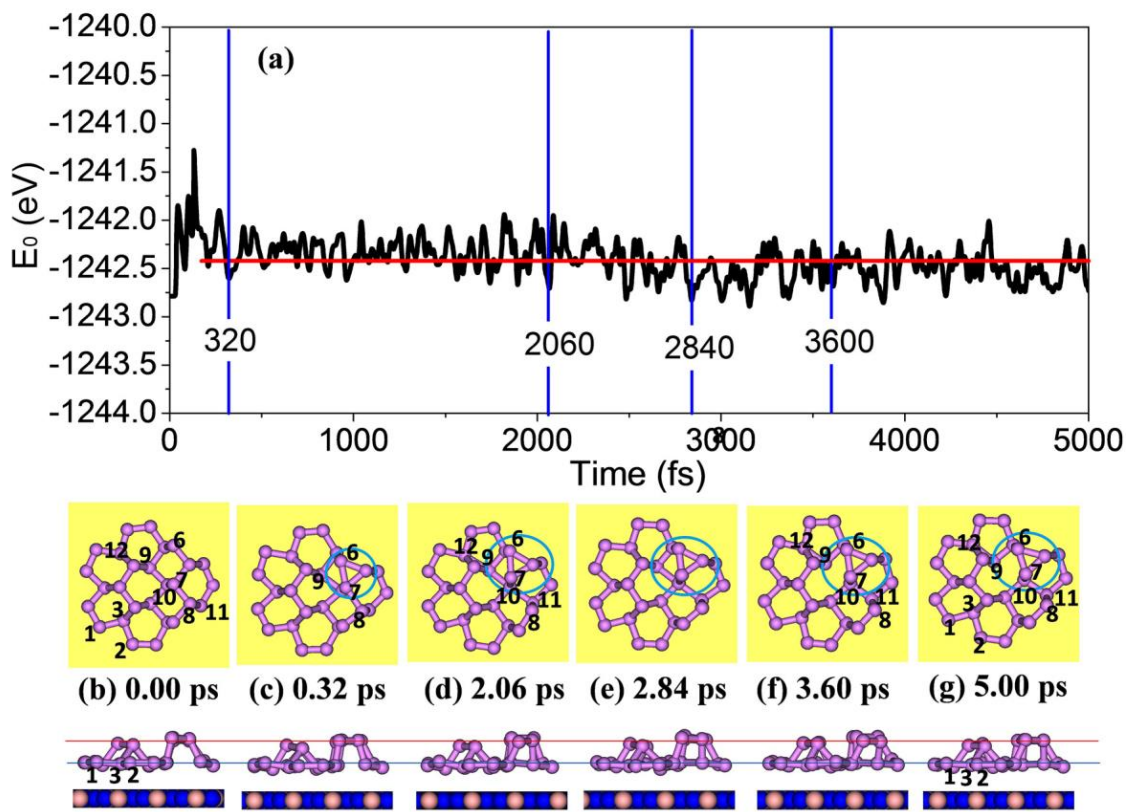

**Figure 4.** The free energy change of the black $P_{27}$ nanoflake on a modified h-BN surface in the 5.0 ps AIMD simulations (a) and the top view and side view of structural evolutions with time (b-g). Blue circles label the structure deformations. The yellow background in top view represents the modified BN substrate for a clear view of $P_{27}$ nanoflake. The lines in side views divide the $P_{27}$ nanoflake in two layers roughly.

**3.4 Moderate substrate interaction to stabilize phosphorene $P_{27}$ nanoflake**

Our above analyses have shown that the Cu(111) substrate is too strong to retain the integrity of the $P_{27}$ structure; while the h-BN substrate is too weak to stabilize the 2D characteristic of the $P_{27}$ nanoflake. For the purposes of direct and simple comparison to the h-BN surface, we proposed a theoretical substrate by artificially enhancing the vdW interaction between the black P nanoflake and h-BN substrate. Hereafter, the artificial substrate is called modified h-BN, whose $C_6$ parameters for B and N species were increased 15 times, and as a result, the interaction energy between $P_{27}$ and h-BN substrate was enhanced to 0.349 eV/P atom (Table 1). Since the value of $C_6$ for P was not changed, the interactions between phosphorus atoms were not affected. Our *ab initio* energy minimization showed that on this artificial substrate, the edge of $P_{27}$ nanoflake was more flat than those in vacuum and on the h-BN surface. In particular, the inner part of the $P_{27}$ nanoflake still retained its two-layer structure (see Figure 4).

The AIMD calculations are shown in Figure 4. It is interesting to see that both the structure and energy of black $P_{27}$ were almost unchanged on the artificially modified h-BN surface, and its two-layer characteristic remained intact. During the whole 5.0 ps AIMD simulation, the P1-P2 bond was never formed and thus the edge had no significant reconstruction (See Figure 4 and Movie-S5 in SI). The P6-P7 bond was formed at 0.32 ps (Figure 4c), and subsequently the P7-P8 bond was broken at ~ 2.06 ps (Figure 4d). At the same time, a new P-P bond was formed between P10 and P11, right beneath the P7 and P8 atoms (Figure 4d&e), respectively. After ~3 ps AIMD simulation, the P9-P10 bond was broken, which transformed the four-coordination P10 atom into three-coordination, but P12 atom in the second edge became a new four-coordination atom (Figure 4f&g). It is noted that all these deformations belong to the intra-layer redistribution without any formation of new layers (see the Movie-S6 in SI). From 0.0 to 5.0 ps, there was negligible change in the average total energy, which is in contrast to the obvious energy reduction observed in the $P_{27}$ nanoflake in vacuum and on the h-BN surface.

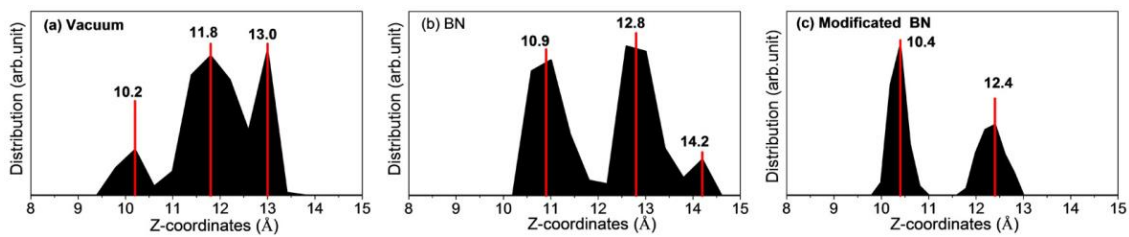

**Figure 5.** The distribution of the inner atoms of the $P_{27}$ nanoflake in the last ~0.5 ps simulations along the z coordinates. (a) in vacuum, (b) on the h-BN surface, (c) on the modified h-BN surface.

### 3.5 Discussion

Figure 5 showed the distributions of the inner atoms in the black $P_{27}$ (three-coordination P atoms in the original structure) along z direction in the last ~0.5 ps of AIMD simulations. It is seen that the two-layer black $P_{27}$ nanoflake was split into three layers both in vacuum and on the h-BN surface (Figure 5a&b). In vacuum, the newborn layer was formed by downwarping the edge atoms and the new layer distances were very close (~1.2Å – 1.6 Å roughly in Figure 5a). On the h-BN surface, the edge downwarping was prevented by the substrate; thereby the newborn layer was formed by lifting the P3-like edge atoms (Figure 3e-i). The distance between bottom and middle P layers was about 1.9 Å, close to the initial layer distance of phosphorene. But the layer distance between middle and newborn layers was only ~1.4 Å, close to the layer distance of the deformed P nanoflake in vacuum (Figure 5a) and the height of freestanding linear P chains.[64] The distribution analyses agreed with the dividing line in Figures 2 and 3 well. Besides the newborn third layer, all the distribution peaks in vacuum and on the h-BN surface overlapped together, indicating an amorphous tendency. In contrast, there were only two distinct peaks for the $P_{27}$ on the artificially modified h-BN surface. The distance is ~2.0 Å, in agreement with the pristine layer distance of phosphorene. The peak of the bottom layer was higher and sharper due to the substrate restriction on the atomic vibration along z direction, while the peak of upper layer is wider and more disperse. In addition, the bottoms of these two peaks were clearly apart, distinctively different from the overlapped bottoms of distribution peaks in vacuum and on the h-BN surface. Such clearly divided peaks signify the two-layer characteristic of phosphorene, indicating the strong stabilizing effect of this artificial h-BN substrate.

To understand the superiority of this artificial substrate over the standard h-BN surface, we have performed structural transformation analyses on the AIMD simulations. As shown in Figures 2&3, the newborn layers were all induced by the edge wrapping, which started from the bonding of P1 and P2 atoms. Therefore, we explored the bonding processes of P1 and P2 atoms on the h-BN surface with different interaction strengths and compared their bonding energy barriers. As shown in Figure 6a, the energy barrier on the standard h-BN surface is only ~17.5 meV, which is even lower than the thermal fluctuation energy of 26 meV ($k_BT$, $k_B$ is the Boltzmann constant) at 300 K. The structure of transition state was shown in Figure 6b&c (blue arrows indicate the moving direction of the P1 and P2 atoms). The average binding time $\tau$ can be roughly estimated by the formula:

$$\tau = (h/k_BT) \times exp(\Delta E/k_BT) \tag{1}$$

where $h$ and $k_B$ are the Plank and Boltzmann constants, respectively. $T$ is the temperature and $\Delta E$ is the bonding energy barrier. Therefore, the average bonding time is only about 0.3 ps on the h-BN surface at 300 K (Figure 6d), in agreement with our AIMD results. Besides, if the temperature is reduced to 50 K, the bonding is still very easy, only within ~10 ps on average (Figure 6d). If we increase three times the $C_6$ parameter for the B and N species in Grimme's approach[61], the bonding energy barrier is slightly lifted to 27.6 meV and the average bonding time is only increased to 0.5 ps at room temperature.

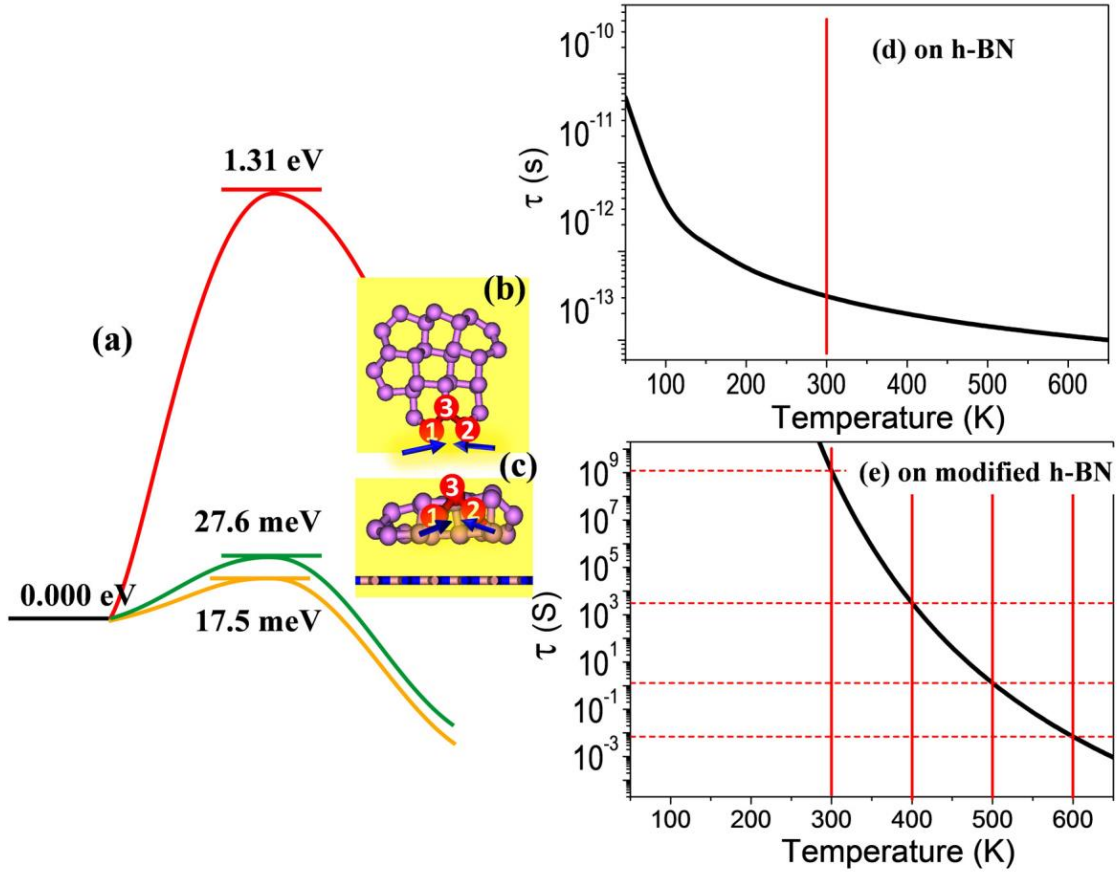

**Figure 6.** The bonding energy barriers (a) of P1 and P2 atoms on the standard h-BN surface (yellow line), the modified h-BN with 3×$C_6$ parameters (green line) and the 15×$C_6$ parameters (red line); the top view (b) and side view (c) of the structure at the saddle point on the h-BN surfaces; the estimated average bonding time for P1-P2 atoms on the standard h-BN (d) and the modified h-BN with 15×$C_6$ parameters (e).

However, If the $C_6$ parameters for the B and N species in Grimme's approach are increased fifteen times, the bonding energy barrier is significantly increased to ~1.31 eV (Figure 6a), which is about 75 times higher than that on the standard h-BN surface. It is worth noting that the energy after P1-P2 bonding is 0.28 eV higher than the original black $P_{27}$ on the modified h-BN surface (Figure S9), indicating that such a deformation is energetically unfavorable, in strong contrast to that on the standard h-BN surface (Figure S7). As a result, on such a moderate interaction surface, the average bonding time for P1 and P2 atoms is dramatically extended to ~$1.2 \times 10^9$ s (longer than years) at 300 K (Figure 6e). That is why there was no edge wrapping of the black $P_{27}$ nanoflake on the artificial h-BN surface during 5.0 ps AIMD simulation. At 400 K, the average bonding time is still $3 \times 10^3$ s (~1 hour) according to Figure 6e, but it drops to 1.0 s at 500 K. When the temperature is elevated up to ~600 K (near the thermal decomposition temperature of phosphorene[72]), the average bonding time for P1 and P2 decreases to only $6.0 \times 10^{-3}$ s. Therefore, to avoid the edge wrapping in the black phosphorene nanoflake, the growth temperature should be carefully controlled during the direct epitaxial growth of phosphorene.

It is worth mentioning that in the present work, the system is placed in vacuum condition. This setting is similar to that of molecular beam epitaxy, in which epitaxial growth is conducted in high vacuum or ultra-high vacuum condition. A similar simulation setting was also adopted previously in many *ab initio* molecular dynamics simulations on the growth of nanomaterials, for example, graphene[75, 76] and nanotube[77].

4. Conclusion

In summary, combining the *ab initio* structure optimizations, molecular dynamics simulations and transition state searching, we investigated the thermal and dynamic stability of phosphorene $P_{27}$ nanoflake on several substrates. The structures of the black $P_{27}$ are strongly dependent on the interaction with substrates. A strong nanoflake-substrate interaction, for example, on Cu(111), will break down the phosphorene nanoflake; while a weak nanoflake-substrate interaction fails to stabilize the phosphorene nanoflake. The instability mainly arises from the edge reconstruction, which transforms the two-layered black phosphorene nanoflake into a three-layer amorphous structure. The energy barrier for edge construction is only 17.5 meV on the h-BN surface, indicating a fast edge reconstruction. On the artificial h-BN substrate with enhanced interaction strength, not only the barrier is lifted up to 1.31 eV, but also the deformation is becoming energetically unfavorable. As a result, the edge wrapping is prohibited and the inner part of the black $P_{27}$ nanoflake is able to retain the 2D characteristic of phosphorene. The binding energy between the black $P_{27}$ nanoflake and the modified h-BN surface is about 0.35 eV/atom, close to that of silicene on the Ag(111) surface.[6, 8] This implies that a substrate with moderate interaction strength is preferred for the epitaxial growth of phosphorene. The next step is to modify the interaction strength so as to satisfy the interaction criterion. Possible ways include selecting suitable substrates, substrate doping, and applying external pressure *etc*.

**Authors**

Junfeng Gao: gaojf@ihpc.a-star.edu.sg;

Gang Zhang: zhangg@ihpc.a-star.edu.sg;

Yong-Wei Zhang: zhangyw@ihpc.a-star.edu.sg

**Acknowledgement**

This work was supported in part by a grant from the Science and Engineering Research Council

(152-70-00017). The authors gratefully acknowledge the financial support from the Agency for Science, Technology and Research (A*STAR), Singapore and the use of computing resources at the A*STAR Computational Resource Centre, Singapore, and Chinese National Supercomputer Center in Tianjin.

**REFERENCES**

1. Tay, R. Y.; Griep, M. H.; Mallick, G.; Tsang, S. H.; Singh, R. S.; Tumlin, T.; Teo, E. H. T.; Karna, S. P. *Nano Lett.* **2014,** 14, 839-846.

2. Song, X.; Gao, J.; Nie, Y.; Gao, T.; Sun, J.; Ma, D.; Li, Q.; Chen, Y.; Jin, C.; Bachmatiuk, A.; Rümmeli, M.; Ding, F.; Zhang, Y.; Liu, Z. *Nano Res.* **2015,** 8, 3164-3176.

3. Gong, Y.; Lei, S.; Ye, G.; Li, B.; He, Y.; Keyshar, K.; Zhang, X.; Wang, Q.; Lou, J.; Liu, Z.; Vajtai, R.; Zhou, W.; Ajayan, P. M. *Nano Lett.* **2015,** 15, 6135-6141.

4. Chhowalla, M.; Liu, Z.; Zhang, H. *Chem. Soc. Rev.* **2015,** 44, 2584-2586.

5. Vogt, P.; De Padova, P.; Quaresima, C.; Avila, J.; Frantzeskakis, E.; Asensio, M. C.; Resta, A.; Ealet, B.; Le Lay, G. *Phys. Rev. Lett.* **2012,** 108, 155501.

6. Gao, J.; Zhao, J. *Sci. Rep.* **2012,** 2, 861.

7. Feng, B.; Ding, Z.; Meng, S.; Yao, Y.; He, X.; Cheng, P.; Chen, L.; Wu, K. *Nano Lett.* **2012,** 12, 3507-3511.

8. Resta, A.; Leoni, T.; Barth, C.; Ranguis, A.; Becker, C.; Bruhn, T.; Vogt, P.; Le Lay, G. *Sci. Rep.* **2013,** 3, 2399.

9. Li, L.; Yu, Y.; Ye, G. J.; Ge, Q.; Ou, X.; Wu, H.; Feng, D.; Chen, X. H.; Zhang, Y. *Nat. Nano.* **2014,** 9, 372-377.

10. Liu, H.; Neal, A. T.; Zhu, Z.; Luo, Z.; Xu, X.; Tománek, D.; Ye, P. D. *ACS Nano* **2014,** 8, 4033-4041.

11. Andres, C.-G.; Leonardo, V.; Elsa, P.; Joshua, O. I.; Narasimha-Acharya, K. L.; Sofya, I. B.; Dirk, J. G.; Michele, B.; Gary, A. S.; Alvarez, J. V.; Henny, W. Z.; Palacios, J. J.; Herre, S. J. v. d. Z. *2D Mater.* **2014,** 1, 025001.

12. Qiao, J.; Kong, X.; Hu, Z.-X.; Yang, F.; Ji, W. *Nat. Commun.* **2014,** 5, 4475.

13. Cai, Y.; Ke, Q.; Zhang, G.; Feng, Y. P.; Shenoy, V. B.; Zhang, Y.-W. *Adv. Func. Mater.* **2015,** 25, 2230-2236.

14. Ong, Z.-Y.; Cai, Y.; Zhang, G.; Zhang, Y.-W. *J. Phys. Chem. C* **2014,** 118, 25272-25277.

15. Fei, R.; Yang, L. *Nano Lett.* **2014,** 14, 2884-2889.

16. Kim, J.; Baik, S. S.; Ryu, S. H.; Sohn, Y.; Park, S.; Park, B.-G.; Denlinger, J.; Yi, Y.; Choi, H. J.; Kim, K. S. *Science* **2015,** 349, 723-726.

17. Umar Farooq, M.; Hashmi, A.; Hong, J. *Sci. Rep.* **2015,** 5, 12482.

18. Tran, V.; Soklaski, R.; Liang, Y.; Yang, L. *Phys. Rev. B* **2014,** 89, 235319.

19. Cai, Y.; Zhang, G.; Zhang, Y.-W. *Sci. Rep.* **2014,** 4, 6677.

20. Das, S.; Zhang, W.; Demarteau, M.; Hoffmann, A.; Dubey, M.; Roelofs, A. *Nano Lett.* **2014,** 14, 5733-5739.

21. Zhang, R.; Li, B.; Yang, J. *J. Phys. Chem. C* **2015,** 119, 2871-2878.

22. Morgan Stewart, H.; Shevlin, S. A.; Catlow, C. R. A.; Guo, Z. X. *Nano Lett.* **2015,** 15, 2006-2010.

23. Fei, R.; Tran, V.; Yang, L. *Phys. Rev. B* **2015,** 91, 195319.

24. Liu, Q.; Zhang, X.; Abdalla, L. B.; Fazzio, A.; Zunger, A. *Nano Lett.* **2015,** 15, 1222-1228.

25. Li, Y.; Wei, Z.; Li, J. *Appl. Phys. Lett.* **2015,** 107, 112103.

26. Liang, L.; Wang, J.; Lin, W.; Sumpter, B. G.; Meunier, V.; Pan, M. *Nano Lett.* **2014,** 14, 6400-6406.

27. Castro Neto, A. H.; Guinea, F.; Peres, N. M. R.; Novoselov, K. S.; Geim, A. K. *Rev. Mod. Phys.* **2009,** 81, 109-162.

28. Palummo, M.; Bernardi, M.; Grossman, J. C. *Nano Lett.* **2015,** 15, 2794-2800.

29. Ling, X.; Wang, H.; Huang, S.; Xia, F.; Dresselhaus, M. S. *Proc. Natl. Acad. Sci. USA.* **2015,** 112, 4523-4530.

30. Chen, Y.; Sun, J.; Gao, J.; Du, F.; Han, Q.; Nie, Y.; Chen, Z.; Bachmatiuk, A.; Priydarshi, M. K.; Ma, D.; Song, X.; Wu, X.; Xiong, C.; Rümmeli, M. H.; Ding, F.; Zhang, Y.; Liu, Z. *Adv. Mater.* **2015,** 27, 7839 - 7846.

31. Cui, Y.; Fu, Q.; Zhang, H.; Bao, X. *Chem. Comm.* **2011,** 47, 1470-1472.

32. Lacovig, P.; Pozzo, M.; Alfè, D.; Vilmercati, P.; Baraldi, A.; Lizzit, S. *Phys. Rev. Lett.* **2009,** 103, 166101.

33. Liu, X.; Fu, L.; Liu, N.; Gao, T.; Zhang, Y.; Liao, L.; Liu, Z. *J. Phys. Chem. C* **2011,** 115, 11976-11982.

34. Reina, A.; Thiele, S.; Jia, X.; Bhaviripudi, S.; Dresselhaus, M.; Schaefer, J.; Kong, J. *Nano Res.* **2009,** 2, 509-516.

35. Liu, K.-K.; Zhang, W.; Lee, Y.-H.; Lin, Y.-C.; Chang, M.-T.; Su, C.-Y.; Chang, C.-S.; Li, H.; Shi, Y.; Zhang, H.; Lai, C.-S.; Li, L.-J. *Nano Lett.* **2012,** 12, 1538-1544.

36. Ling, X.; Lee, Y.-H.; Lin, Y.; Fang, W.; Yu, L.; Dresselhaus, M. S.; Kong, J. *Nano Lett.* **2014,** 14, 464-472.

37. Yu, Q.; Jauregui, L. A.; Wu, W.; Colby, R.; Tian, J.; Su, Z.; Cao, H.; Liu, Z.; Pandey, D.; Wei, D.; Chung, T. F.; Peng, P.; Guisinger, N. P.; Stach, E. A.; Bao, J.; Pei, S.-S.; Chen, Y. P. *Nat. Mater.* **2011,** 10, 443-449.

38. Hanlon, D.; Backes, C.; Doherty, E.; Cucinotta, C. S.; Berner, N. C.; Boland, C.; Lee, K.; Harvey, A.; Lynch, P.; Gholamvand, Z.; Zhang, S.; Wang, K.; Moynihan, G.; Pokle, A.; Ramasse, Q. M.; McEvoy, N.; Blau, W. J.; Wang, J.; Abellan, G.; Hauke, F.; Hirsch, A.; Sanvito, S.; O'Regan, D. D.; Duesberg, G. S.; Nicolosi, V.; Coleman, J. N. *Nat. Commun.* **2015,** 6, 8563.

39. Brent, J. R.; Savjani, N.; Lewis, E. A.; Haigh, S. J.; Lewis, D. J.; O'Brien, P. *Chem. Comm.* **2014,** 50, 13338-13341.

40. Gao, J.; Yip, J.; Zhao, J.; Yakobson, B. I.; Ding, F. *J. Am. Chem. Soc.* **2011,** 133, 5009-5015.

41. Gao, J.; Ding, F. *Angew. Chem. Int. Ed.* **2014,** 53, 14031-14035.

42. Chen, W.; Chen, H.; Lan, H.; Cui, P.; Schulze, T. P.; Zhu, W.; Zhang, Z. *Phys. Rev. Lett.* **2012,** 109, 265507.

43. Cui, Y.; Gao, J.; Jin, L.; Zhao, J.; Tan, D.; Fu, Q.; Bao, X. *Nano Res.* **2012,** 5, 352-360.

44. Wang, B.; Ma, X.; Caffio, M.; Schaub, R.; Li, W.-X. *Nano Lett.* **2011,** 11, 424-430.

45. Markov, I. V., *Crystal Growth for Beginners: Fundamentals of Nucleation, Crystal Growth and Epitaxy*. 2nd ed. ed.; World Scientific Publishing Co. Pte. Ltd.: Singapore, 2003. p99

46. Ramasubramaniam, A.; Muniz, A. R. *Phys. Rev. B* **2014,** 90, 085424.

47. Li, W.; Zhang, G.; Zhang, Y.-W. *J. Phys. Chem. C* **2014,** 118, 22368-22372.

48. Gao, J.; Zhao, J.; Ding, F. *J. Am. Chem. Soc.* **2012,** 134, 6204-6209.

49. Artyukhov, V. I.; Hao, Y.; Ruoff, R. S.; Yakobson, B. I. *Phys. Rev. Lett.* **2015,** 114, 115502.

50. Artyukhov, V. I.; Liu, Y.; Yakobson, B. I. *Proc. Natl. Acad. Sci. USA*. **2012,** 109, 15136-15140.

51. Meca, E.; Lowengrub, J.; Kim, H.; Mattevi, C.; Shenoy, V. B. *Nano Lett.* **2013,** 13, 5692-5697.

52. Hao, Y.; Bharathi, M. S.; Wang, L.; Liu, Y.; Chen, H.; Nie, S.; Wang, X.; Chou, H.; Tan, C.; Fallahazad, B.; Ramanarayan, H.; Magnuson, C. W.; Tutuc, E.; Yakobson, B. I.; McCarty, K. F.; Zhang, Y.-W.; Kim, P.; Hone, J.; Colombo, L.; Ruoff, R. S. *Science* **2013,** 342, 720-723.

53. Whitesides, R.; Frenklach, M. *J. Phys. Chem. A* **2010,** 114, 689-703.

54. Ming, F.; Zangwill, A. *Phys. Rev. B* **2011,** 84, 115459.

55. Meng, L.; Sun, Q.; Wang, J.; Ding, F. *J. Phys. Chem. C* **2012,** 116, 6097-6102.

56. Zhao, Y.; Truhlar, D. G. *Accounts of Chemical Research* **2008,** 41, 157-167.

57. Wu, P.; Jiang, H.; Zhang, W.; Li, Z.; Hou, Z.; Yang, J. *J. Am. Chem. Soc.* **2012,** 134, 6045-6051.

58. Kresse, G.; Furthmüller, J. *Phys. Rev. B* **1996,** 54, 11169-11186.

59. Kresse, G.; Furthmüller, J. *Comput. Mater. Sci.* **1996,** 6, 15-50.

60. Perdew, J. P.; Burke, K.; Ernzerhof, M. *Phys. Rev. Lett.* **1996,** 77, 3865-3868.

61. Grimme, S. *J. Comput. Chem.* **2006,** 27, 1787-1799.

62. Blöchl, P. E. *Phys. Rev. B* **1994,** 50, 17953-17979.

63. Henkelman, G.; Uberuaga, B. P.; Jónsson, H. *J. Chem. Phys.* **2000,** 113, 9901-9904.

64. Karttunen, A. J.; Linnolahti, M.; Pakkanen, T. A. *Chem. Eur. J.* **2007,** 13, 5232-5237.

65. Gao, J.; Yuan, Q.; Hu, H.; Zhao, J.; Ding, F. *J. Phys. Chem. C* **2011,** 115, 17695-17703.

66. Avsar, A.; Vera-Marun, I. J.; Tan, J. Y.; Watanabe, K.; Taniguchi, T.; Castro Neto, A. H.; Özyilmaz, B. *ACS Nano* **2015,** 9, 4138-4145.

67. Chen, X.; Wu, Y.; Wu, Z.; Han, Y.; Xu, S.; Wang, L.; Ye, W.; Han, T.; He, Y.; Cai, Y.; Wang, N. *Nat. Commun.* **2015,** 6, 7315.

68. Mishchenko, A.; Cao, Y.; Yu, G. L.; Woods, C. R.; Gorbachev, R. V.; Novoselov, K. S.; Geim, A. K.; Levitov, L. S. *Nano Lett.* **2015,** 15, 6991-6995.

69. Nathaniel, G.; Darshana, W.; Yanmeng, S.; Tim, E.; Jiawei, Y.; Jin, H.; Jiang, W.; Xue, L.; Zhiqiang, M.; Kenji, W.; Takashi, T.; Marc, B.; Yafis, B.; Roger, K. L.; Chun Ning, L. *2D Mater.* **2015,** 2, 011001.

70. Rivero, P.; Horvath, C. M.; Zhu, Z.; Guan, J.; Tománek, D.; Barraza-Lopez, S. *Phys. Rev. B* **2015,** 91, 115413.

71. Hu, T.; Hong, J. *ACS Appl. Mater. Interfaces* **2015,** 7, 23489-23495.

72. Liu, X.; Wood, J. D.; Chen, K.-S.; Cho, E.; Hersam, M. C. *J. Phys. Chem. Lett.* **2015,** 6, 773-778.

73. Nosé, S. *J. Chem. Phys.* **1984,** 81, 511-519.

74. Jones, R. O.; Ganteför, G.; Hunsicker, S.; Pieperhoff, P. *J. Chem. Phys.* **1995,** 103, 9549-9562.

75. Wang, Y.; Page, A. J.; Nishimoto, Y.; Qian, H.-J.; Morokuma, K. Irle, S. *J. Am. Chem. Soc.* **2011,** 133, 18837

76. Jiao, M.; Song, W.; Qian, H.-J.; Wang, Y.; Wu, Z.; Irle, S.; Morokuma, K. *Nanoscale,* **2016,** 8, 3067-3074

77. Raty, J.-Y.; Gygi, F.; Galli, G. *Phys. Rev. Lett.* **2005,** 95, 096103